\documentclass[10pt,twocolumn]{IEEEtran}
\usepackage{graphicx}
\usepackage{amsmath}
\usepackage{amssymb}
\usepackage{cite}
\usepackage{float}
\usepackage{color}
\usepackage{url}
\usepackage{graphicx}
\usepackage{amsfonts}
\usepackage{float}
\usepackage{color}
\usepackage{cite}
\usepackage{epstopdf}
\usepackage{bm}
\usepackage{mathtools}
\usepackage[algoruled,resetcount,linesnumbered]{algorithm2e}
\usepackage{array}
\usepackage{multirow}
\usepackage{multicol}
\usepackage{mathtools}

\newtheorem{remark}{Remark}

\usepackage{lipsum}
\newcommand*{\Scale}[2][4]{\scalebox{#1}{\ensuremath{#2}}}%
\newcommand\blfootnote[1]{%
  \begingroup
  \renewcommand\thefootnote{}\footnote{#1}%
  \addtocounter{footnote}{-1}%
  \endgroup
}

\graphicspath{ {fig/} } 

\begin{document}

\title{Off-Grid Aware Spatial Covariance Estimation in mmWave Communications}
\author{Chethan Kumar Anjinappa,
Ali Cafer G\"{u}rb\"{u}z$^\dagger$, Yavuz Yap{\i}c{\i}, and \.{I}smail G\"{u}ven\c{c} \\
\IEEEauthorblockA{Department of Electrical and Computer Engineering, 
North Carolina State University, Raleigh, NC
}\\
\IEEEauthorblockA{$^\dagger$Department of Electrical and Computer Engineering, Mississippi State University, Mississippi State, MS 
}\\
Email: {\tt canjina@ncsu.edu, gurbuz@ece.msstate.edu, \{yyapici, iguvenc\}@ncsu.edu}\vspace{-.2cm}}

\maketitle

\begin{abstract}
This work investigates the problem of spatial covariance matrix estimation in a millimeter Wave (mmWave) hybrid multiple-input multiple-output (MIMO)  system with an emphasis on the basis-mismatch effect. The basis mismatch is prevalent in the compressed sensing (CS) schemes which adopt discretization procedure. In such an approach, the algorithm yields a finite \textit{discrete} point which is an approximation to the \textit{continuous} parametric space. The quality of this approximation depends on the number of discretized points in the dictionary. Instead of increasing the number of discretized points to combat this \textit{off-grid} effect, we propose an efficient parameter perturbed framework which uses a controlled perturbation mechanism in conjunction with the orthogonal matching pursuit (OMP) algorithm. Numerical results verify the performance improvement through our proposed algorithm in terms of \textit{relative efficiency metric}, which is basically due to taking care of the off-grid effect carefully that is ignored in the conventional CS algorithms.
\blfootnote{This research was supported in part by NSF under the grant numbers ACI-1541108 and ECCS-1611112.}
\end{abstract}

\begin{IEEEkeywords}
Basis mismatch, compressed sensing, covariance estimation, 
mmWave covariance estimation, off-grid, orthogonal matching pursuit, parameter perturbed framework.
\end{IEEEkeywords}

\section{Introduction}
The multiple-input multiple-output (MIMO) is a quintessential technology for 5G millimeter Wave (mmWave) communications to deliver higher data rates, higher spectral efficiency, and lower latency, exceeding what is possible with the traditional cellular systems. At mmWave bands, the \textit{hybrid analog/digital beamforming (HADB)} architecture is adapted \cite{mendez2016hybrid,heath2016overview} to curtail the issues of the conventional MIMO mmWave architecture. One of the critical challenges with the HADB architecture is to configure the analog precoding stages. In most of the prior works, this is accomplished by assuming the availability of full channel state information (CSI) at the transmitter, which is difficult to obtain even for a time division duplexing system. As a promising alternative, a \textit{spatial covariance matrix} based method has recently been proposed to update the analog RF precoders, which does not need full CSI and exploits the relatively stationary long-term channel statistics \cite{li2017optimizing,park2018spatial}. 

Based on the way the spatial covariance matrix is estimated, it can be broadly categorized into two methods: 1) \textit{covariance estimation via the channel estimation framework} which we refer to as the indirect method, and 2) \textit{explicit covariance estimation} referred as the direct method. The central idea in the indirect approach is to solve for the channel estimates for every successive snapshot and use these estimates to calculate the covariance matrix. Upon obtaining the channel estimates for every snapshot, the covariance calculation is relatively straightforward. However, in cases when the channel estimates are not required, then one can explicitly operate on the covariance of measurements directly to estimate the covariance matrix which is central to the latter approach. This single step of directly recovering the second order statistics allows for a better exploitation of the statistical structure. Whereas, the former approach incurs more computational complexity and limits the compression ratio \cite{romero2015compressive}.

Both the \textit{channel estimation} and the \textit{covariance estimation} problems can be posed as a compressed sensing (CS) problem leveraging the sparse nature of mmWave channel \cite{ChannelModel_Rappaport}. There have been various smart algorithms leveraging the virtual channel model along with CS techniques \cite{park2018spatial,romero2015compressive,ChannelModel_Rappaport,Vector_Type}. These algorithms, however, either assume that the virtual channel model is exactly sparse, or increase the number of discretized points in the dictionary to end up with sparser structure, which impairs their efficiency \cite{2018_POMP_ICC, chi2011sensitivity}. In fact, for the discrete Fourier transform (DFT) basis defined by the virtual channel model, a continuous parameter lying between two successive DFT grid cells will affect not only the closest two cells, but the whole grid with amplitude decaying with $\frac{1}{N}$ due to the \textit{Dirichlet kernel} \cite{teke2013perturbed,Dirichlet_Kernel}, where $N$ is the number of DFT basis points. This \textit{off-grid} phenomena violates the sparsity assumption, resulting in a decrease in reconstruction performance. As a result, the estimation accuracy of the CS based methods is limited by the number of grid points. More details on the basis mismatch/off grid effects can be found in \cite{Dirichlet_Kernel,teke2013perturbed,teke2014robust,tang2013compressed_off_grid,chi2011sensitivity}.

In this work, we propose a \textit{parameter perturbed framework} to combat the basis-mismatch problem in spatial covariance estimation. The proposed algorithm evade the issue arising from the basis mismatch problems by operating on the continuum angle-of-arrival (AoA) and angle-of-departure (AoD) space using the
mechanism of the controlled perturbation in conjunction with a modified orthogonal matching pursuit (OMP) framework. The key in the designed parameter-perturbed framework is to preserve the sub-optimal greedy projection step of the OMP algorithm and then invoke controlled perturbation mechanism on the selected columns from the projection step. This procedure allows one to combat the off-grid effects after the projection step and before the update of the residual terms
which is the central innovation behind the developed parameter perturbed algorithm. We present the rationale behind this central innovation and validate the superiority of the proposed
method by numerical simulations.

{\bf Notation}: Vectors and matrices are represented by lower-case (eg: \textbf{a}) and upper-case boldface (eg: \textbf{A}) letters, respectively. Every vector is considered as a column vector. The transpose, conjugate, conjugate transpose, and pseudo-inverse of a matrix \textbf{A} are denoted by  $\textbf{A}^\text{T}$, $\textbf{A}^\text{H}$, $\textbf{A}^{*}$, and $\textbf{A}^{\dagger}$, respectively. For an integer $K$, we use the shorthand notation $[K]$ for the set of non-negative integers $\{1,2,\ldots, K\}$.

\section{System Model and Channel Model}
Consider a mmWave MIMO network comprised of a base station (BS) communicating with generic user equipment (UE), both equipped with a uniform linear array. We assume that the BS is equipped with $M$ antennas and $M_{\rm RF}$ radio frequency (RF) chains. Similarly, the UE is equipped with $N$ antennas and $N_{\rm RF}$ RF chains. Typically, it is assumed that $M_{\rm RF}\leq M$ and $N_{\rm RF}\leq N$. With HADB MIMO processing structure, the received signal at the UE per snapshot can be expressed as
\begin{eqnarray}\label{SSR_Setup}
    \mathbf{y}_t =  \mathbf{\Phi} {\mathbf{h}_t} + \mathbf{W} \mathbf{n}_t, \quad \forall t \in [T],
\end{eqnarray}
where $\mathbf{\Phi}= ({\mathbf{f}^\text{T} \otimes \mathbf{W}})$ is the combined effect of the precoder/combiner; $\mathbf{W}$ and $\mathbf{f}$ represent the combiner matrices and precoder vectors, respectively, and it is assumed to be time-invariant across the frame. The noise vector at the UE, $\mathbf{n}_t$, follows a circularly symmetric independent and identically distributed Gaussian distribution, $\mathcal{CN}(0,\sigma_n^2)$ with noise variance $\sigma_n^2$. The term $\mathbf{h}_{t}$\footnote{Rewritten using the vector identity property: vec($\bf{ABC}$) = $(\mathbf{C}^T \otimes \mathbf{A}) \text{vec}(\mathbf{B})$; $\otimes$ denotes the Kronecker product.} denote the vectorized version of the original channel matrix $\mathbf{H}_t$ at time-frame~$t$.

Following the model in \cite{ChannelModel_Rappaport}, we assume the original channel $\mathbf{H}_t$ to be composed of $K$ spatial path clusters with each cluster containing $L$ macro-level scattering multi-path components (MPCs) and is represented as: 
\begin{equation}\label{H_Equation}
\mathbf{H}_t = \frac{1}{\beta}\sum_{k=1}^K \sum_{l=1}^L \alpha_{k,l,t} \mathbf{a}_\text{UE}(\theta_{k,l}^\text{rx}) \mathbf{a}_\text{BS}(\theta_{k,l}^\text{tx})^\text{H}, \quad \forall t \in [T] 
\end{equation}
Note that $K$ and $L$ may each be time-varying due to mobility of the UE and the surrounding scatterers \cite{Angular_Temporal_Correlation_Chethan_2018,Propogation_Statistics_Chethan_2018}. However, for simplicity, we assume $K$ and $L$ to be fixed at least for the duration of the covariance estimation. Here on, we use the short notation of $K_L$ to represent a total of $KL$ MPCs. The term $\alpha_{k,l,t}$ and $\beta$ denotes the small scale fading time-varying complex gain of the $\{k,l\}^\text{th}$ MPC during the $t^\text{th}$ frame and the normalization factor, respectively. The terms $\theta_{k,l}^\text{rx}$ and $\theta_{k,l}^\text{tx}$ denote the azimuthal AoA and AoD of the $\{k,l\}^\text{th}$ MPC. Unlike $\alpha_{{k,l},t}$, the AoA and AoD are assumed to be constant across the $T$ snapshots. This assumption is due to the fact that the angular coherence time is much larger than the channel coherence
time \cite{Beamwidth_Temporal_Channel_Variataion_Heath}.
The function $\mathbf{a}_\text{BS}(\theta_{i}^\text{tx})$ represent the BS array response vector to the AoD $\theta_{i}^\text{tx}$ and is expressed as $[\mathbf{a}_\text{BS} (\theta_{i}^\text{tx})]_m = \frac{1}{\sqrt{M}} e^{j \pi(m-1) \cos(\theta_{i}^\text{tx})}, \quad m \in [M]$. Similarly,  $\mathbf{a}_\text{UE}(\theta_{i}^\text{rx})$ which represent the UE array response.

In order to apply the CS schemes, researchers typically adopt a discretization (or gridding) procedure which reduces the continuous parameter space, continuum AoA-AoD tuple ($\theta^\text{rx} , \theta^\text{tx}$) in the interval $\left([ 0, \pi ) \times [ 0, \pi )\right)$, into a set of finite grid points. To denote it mathematically, we consider $\Theta^\text{rx}$ and $\Theta^\text{tx}$ as the set containing the $G_\text{UE}$ and $G_\text{BS}$ finite discretized  grid points in the AoA and AoD domains, respectively. These discretized grid points are chosen such that they satisfy certain properties based on the scheme adapted. The two popular schemes include uniform sampling of the physical and virtual domains, respectively. In this work, we adopt the latter approach \cite{Uniform_Sampling_Cos_Domain} which reduces the coherence of the redundant dictionary due to preserving orthogonality which does not hold in the former approach.

In the uniform sampling of the virtual domain, the AoA/AoD are taken from a non-uniformly quantized grid such that the $\left(\cos(\theta^\text{rx}), \cos(\theta^\text{tx})\right)$ space appearing in the array response is uniformly quantized \cite{Uniform_Sampling_Cos_Domain,2017_Robust_Beam_Alignment}. The grid angles in this approach follow the inverse cosine function as follows
\begin{eqnarray*}
\begin{aligned}
\Theta^\text{tx} &= \Scale[1.05]{\left\{\bar{\theta}^\text{tx}_{i}: \cos(\bar{\theta}^\text{tx}_{i}) = 1 - \frac{2(i-1)}{G_\text{BS}} \in [1,-1), i \in [G_\text{BS}]\right\}},\\
\Theta^\text{rx} &= \Scale[1.05]{\left\{\bar{\theta}^\text{rx}_{i}: \cos(\bar{\theta}^\text{rx}_{i}) = 1 - \frac{2(i-1)}{G_\text{UE}} \in [1,-1), i \in [G_\text{UE}]\right\}}.
\end{aligned}
\end{eqnarray*}
Collecting all the array responses corresponding to the angles from the set ($\Theta^\text{rx} , \Theta^\text{tx} $), the array response matrices $\mathbf{A}_\text{UE}(\Theta^\text{rx}) = [\mathbf{a}_\text{UE}(\bar{\theta}_{1}^\text{rx}), \ldots , \mathbf{a}_\text{UE}(\bar{\theta}_{G_\text{UE}}^\text{rx}) ]$ and $\mathbf{A}_\text{BS}(\Theta^\text{tx}) = [\mathbf{a}_\text{BS}(\bar{\theta}_{1}^\text{tx}), \ldots, \mathbf{a}_\text{BS}(\bar{\theta}_{G_\text{BS}}^\text{tx}) ]$ are formed. Using these array response matrices, the channel matrix $\mathbf{H}_t$ can be represented by the virtual sparse channel which provides a discretized approximation of the channel response
\begin{equation}\label{Virtual_Channel}
    \mathbf{H}_t = \mathbf{A}_\text{UE} \mathbf{H}_{V_t} \mathbf{A}_\text{BS}^\text{H},  \quad \forall t \in [T],
\end{equation}
where $\mathbf{H}_{V_t} \in \mathcal{C}^{G_\text{UE} \times G_\text{BS}}$ is the sparse virtual matrix containing the quantized spatial frequencies. Aided by the sparse virtual representation and vector identity property, the MIMO channel estimation (\ref{SSR_Setup}) is rewritten as a sparse recovery problem
\begin{eqnarray}\label{SSR_Problem}
    \mathbf{y}_t =  \mathbf{\Phi} \mathbf{\Psi} {\mathbf{h}_{v_t}} + \mathbf{W} \mathbf{n}_t, \quad \forall t \in [T],
\end{eqnarray}
where $\mathbf{h}_{V_t} = \text{vec}(\mathbf{H}_{V_t})$ is the vectorized version of the sparse virtual matrix $\mathbf{H}_{V_t}$. The term $\mathbf{\Psi} = (\mathbf{A}_\text{BS}^\text{C} \otimes \mathbf{A}_\text{UE})$ is the combined effect of the dictionary matrix containing the array responses corresponding to the discretized spatial angles. The conventional CS techniques assume that the virtual channel $\mathbf{h}_{V_t}$ is exactly sparse, which is true only when the AoA-AoD tuples are aligned with discretized spatial angles which is an ideal on-grid case. However, the physical AoA-AoD can take any continuous values as defined in model (\ref{H_Equation}), which may not be aligned with any discretized spatial angles causing  off-the-grid effects. These effects violate the sparsity assumption, resulting in performance degradation of CS-based techniques \cite{2018_POMP_ICC, chi2011sensitivity,Dirichlet_Kernel,teke2013perturbed}. 

The goal of this work is to estimate the channel sample covariance matrix $\mathbf{{R}}_h =\mathop{\mathbb{E}}(\mathbf{h}_{t} \mathbf{h}_{t}^\text{H})$ using the finite $T$ under-determined set of measurements $\mathbf{{y}}_{t}$ considering the off-grid effects. Next we discuss this off-grid phenomena and provide an off-grid aware representation in conjunction with the discretized dictionary.


\section{Capturing Off-the-Grid Effects}
The source of the off-grid complication is that the true continuum AoA/AoD tuple $({\theta}^\text{rx},{\theta}^\text{tx})$ may not coincide with one of the predefined discretized grid points in $(\Theta^\text{rx},\Theta^\text{tx})$, but may be perturbed away from these grid points with unknown perturbation. This implies that the $\mathbf{h}_{V_t}$ may not be exactly sparse in the assumed basis $\mathbf{\Psi}$, but in the unknown basis $\mathbf{\hat{\Psi}}$. Since it is assumed that the total number of MPCs is $K_L$, there exist $K_L$ columns of $\mathbf{\Psi}$ that needs to be updated adaptively. We denote the indices corresponding to these $K_L$ columns as $\mathcal{K_L}$. At first, we investigate the perturbation mechanism for a single MPC. After we see how to address a single MPC, extending it to accommodate multiple MPCs is relatively straightforward.

%

Mathematically, the true AoA $\theta_l^\text{rx}$ of the $l^\text{th}$ MPC and the perturbation from the nearest grid point can be related as: $\theta_l^\text{rx} = \bar{\theta}_l^\text{rx} + \delta_l^\text{rx}$, where $\bar{\theta}_l^\text{rx}$ is the grid point that is closest to the true AoA from the set $\Theta^\text{rx}$, and $\delta_l^\text{rx}$ is the perturbation parameter in a bounded AoA space. This bounded space is dependent on the sampling scheme and the number of grid points employed during the creation of a dictionary matrix.  A similar relation holds for the true AoD and the AoD perturbation as $\theta_l^\text{rx} = \bar{\theta}_l^\text{rx} + \delta_l^\text{rx}$. The unknown basis for the $l^\text{th}$ MPC can then be related to the nearest discretized grid and perturbation as follows
\begin{align}
{\hat{\Psi}}_l &= \text{vec}\left(\mathbf{a}_\text{UE}(\theta_l^\text{rx})\mathbf{a}_\text{BS}(\theta_l^\text{tx})^\text{H}\right) 
\\ &= \text{vec}\left(\mathbf{a}_\text{UE}(\bar{\theta}_l^\text{rx} + \delta_l^\text{rx})\mathbf{a}_\text{BS}(\bar{\theta}_l^\text{tx} + \delta_l^\text{tx})^\text{H}\right). 
\end{align}
The unknown basis for all the $K_L$ MPCs can be related as $\mathbf{\hat{\Psi}}_\mathcal{K_L} = [ {\Psi}_{1}, \ldots, {{\Psi}}_{K_L} ] $. If the perturbation parameters can be found then the degradation due to off-grid can be reduced significantly. From this perspective, it becomes clear why capturing the perturbations might be necessary for the optimal sparse representation of the virtual channel model. Thus, the key idea is to solve for the perturbations ($\delta_l^\text{tx},\delta_l^\text{rx}$) from the discretized grid points.

\section{Parameter-Perturbed Covariance OMP (PPCOMP) for Covariance Estimation}
 In this work, we explicitly solve for the covariance matrix, as opposed to the method of obtaining the channel estimates from (\ref{SSR_Setup}) and then constructing the covariance matrix. This can be made possible by relating the covariance of the measurements $\mathbf{R}_{y}$ and the virtual covariance matrix $\mathbf{R}_{h_{V}} $ as follows
\begin{eqnarray}\label{Reformulated_Covariance_Problem}
\begin{aligned}
\mathbf{R}_{y} &= \mathop{\mathbb{E}}(\mathbf{y}_{t} \mathbf{y}_{t}^\text{H}) = \mathbf{\Phi} \underbrace{\mathbf{\Psi} \mathbf{R}_{h_{V}}  \mathbf{\Psi}^\text{H}
}_{\approx \mathbf{R}_h} \mathbf{\Phi}^\text{H} +   \bf{E},
\end{aligned}
\end{eqnarray}
where the term $\mathbf{E}$ captures the noise and the signal-noise cross terms of (\ref{SSR_Setup}), and is treated as noise all together. The term $\mathbf{R}_h$ can be explicitly represented as follows: $\mathbf{R}_h = \mathbb{E}(\mathbf{h}_t \mathbf{h}_t^\text{H}) = \sum_{l}^{K_L}\sum_{q}^{K_L} \Gamma_{l,q} \mathbf{A}(\Theta_l,\Theta_q)$, where $\Gamma_{l,q} = \sum_{t=1}^T \alpha_{l,t}\alpha_{q,t}^*$ and $\Theta_l = (\theta_l^\text{rx} ,\theta_l^\text{tx})$ contains both the AoA-AoD tuple. The term $\mathbf{A}(\Theta_l,\Theta_q) = \text{vec}(\mathbf{a}_\text{R}(\theta_{l}^\text{rx}) \mathbf{a}_\text{T}(\theta_{l}^\text{tx})^\text{H})\text{vec}(\mathbf{a}_\text{R}(\theta_{q}^\text{rx}) \mathbf{a}_\text{T}(\theta_{q}^\text{tx})^\text{H})^\text{H}$. Here onward, we use the shorthand notation $\mathbf{a}_\text{res}(\theta_{l}^\text{rx},\theta_{l}^\text{tx}) = \text{vec}(\mathbf{a}_\text{R}(\theta_{l}^\text{rx}) \mathbf{a}_\text{T}(\theta_{l}^\text{tx})^\text{H})$. As discussed, the actual angle parameters and the discretized spatial angles of the $l^\text{th}$ MPC can be related by the perturbation as $\theta_l^\text{rx} = \Bar{\theta}_l^\text{rx} + \delta_l^\text{rx}$, where the perturbation is bounded as $\Delta_{\text{LB}_l} \leq \delta_l^\text{rx}\leq \Delta_{\text{UB}_l}$. The terms $\Delta^\text{rx}_{\text{LB}_l}$ and $\Delta^\text{rx}_{\text{UB}_l}$ are the lower and upper bound for the perturbation in the AoA space for the $l^\text{th}$ MPC. Similarly for the AoD, $\theta_l^\text{tx} =  \Bar{\theta}_l^\text{tx} + \delta_l^\text{tx}$. 

For the uniform sampling of $\cos(\theta)$ scheme, the bounded space for perturbations is non-uniform and is dependent on the nearest grid point in the dictionary. The lower and
upper bound for the perturbation in the AoD space can then be related as $\Delta_\text{LB}^\text{tx} = (\bar{\theta}_{l}^\text{tx} - \bar{\theta}_{l-1}^\text{tx})/2 $ and $\Delta_\text{UB}^\text{tx} = (\bar{\theta}_{l+1}^\text{tx} - \bar{\theta}_{l}^\text{tx})/2$, where $\bar{\theta}^\text{tx}_{l-1}$ and $\bar{\theta}^\text{tx}_{l+1}$ are the adjacent grid points for the chosen initial grid point, respectively. Similarly, $\Delta^\text{rx}_\text{LB}$ and $\Delta^\text{rx}_\text{UB}$ for the AoA space.

With this notation, our goal is to perturb the grid parameters $\bm{\delta}^\text{rx} = \delta_l^\text{rx}$ $\forall l \in [K_L]$, $\bm{\delta}^\text{tx} = \delta_l^\text{tx}$ $\forall l \in [K_L]$, and find $\bm{\Gamma} = \Gamma_{l,q},  \forall l, q \in [K_L]$ in order to minimize the residual error in a noisy environment subject to bounded perturbations. The optimization problem is represented as follows: 
\begin{eqnarray}\label{PPOMP_DCOMP_Opt_Problem}
\begin{aligned}
 \underset{\bm{\Gamma}, \bm{\delta}^\text{rx},\bm{\delta}^\text{tx} }{\text{min}} & \quad  \Bigg| \Bigg|\mathbf{R}_{y}- \mathbf{\Phi} \Bigg( \sum_{l,q}^{K_L}\Gamma_{l,q} \mathbf{A}(\Theta_l,\theta_q) \Bigg) \mathbf{\Phi}^\text{H}\Bigg|\Bigg|_F^2,
 \\ \textit{s.t.} &\quad \Delta^\text{tx}_{\text{LB}_l}  \leq \delta^\text{tx}_l \leq  \Delta^\text{tx}_{\text{UB}_l}, \quad  \Delta^\text{rx}_{\text{LB}_l} \leq \delta^\text{rx}_l \leq  \Delta^\text{rx}_{\text{UB}_l}.
\end{aligned}
\end{eqnarray}
The above joint optimization problem (\ref{PPOMP_DCOMP_Opt_Problem}) is non-convex and in general challenging to solve. We solve the above problem (\ref{PPOMP_DCOMP_Opt_Problem}) in a greedy iterative fashion, where we split the problem into finding the initial grid points for each MPC and proposing a perturbation mechanism for perturbing the MPCs.

\begin{algorithm}[b!]
\KwIn{${{\mathbf{y}}_{t}} \forall t \in [T]$, ${{\Phi}}$, $\Psi$, $\epsilon$
\\ \textbf{Initialization:} ${\mathbf{R}_{{y}_{{\perp}}}} = \mathbf{R}_{y}$, ${\mathcal{S}_{0}} = \{ \}$, $ e  =  || \mathbf{R}_{y}||_F^2$, $k$ = 1.}
\While{$e < \epsilon$}{
{
$j^{\star} = \arg \underset{j}{\max} \quad | ({\Phi}{\Psi})_j^\text{H} {\mathbf{R}_{{y}_{{\perp}}}}({\Phi}{\Psi})_j|$\\
$\mathcal{S}_{k} = \mathcal{S}_{k-1} \cup j^\star$\\
$({\bm{\Gamma}, \bm{\delta}^\text{rx}, \bm{\delta}^\text{rx}} ) = \mathbb{S}(\mathbf{R}_y, \mathcal{S}_k)$\\
$\mathbf{R}_{y_{\perp}} = \mathbf{R}_{y} - \sum_{l,q}^{K_L}\Gamma_{l,q} \mathbf{A}(\Theta_l,\theta_q)$\\
$e$ = $|| \mathbf{R}_{y_{\perp}}||_F^2$\\
$k = k + 1$
}
}
\KwOut{$\mathbf{R}_{h} $}
\caption{Covariance Estimation: PPCOMP} \label{Algorithm:PDCOMP}
\end{algorithm}

\subsection{Finding Initial Grid Points}
The goal at each iteration is to choose an initial grid point which minimizes the orthogonal residual as much as possible and this is achieved by the classical projection {Step~2} in Algorithm 1. The notable change in the projection step compared to the standard OMP is the use of quadratic forms instead of the linear forms to accommodate the measurement covariance \cite{park2018spatial}. The first implication is that the index $j^\star$ chosen by the projection step indicates the discretized point most correlated to the true AoA-AoD tuple among all the possible discretized AoA/AoD tuple. Intuitively, this step provides the initial grid points ($\bar{\theta}^\text{rx}_l,\bar{\theta}^\text{tx}_l$) from the predefined discretized set ($\Theta^\text{rx},\Theta^\text{tx}$). The second implication is that this allows one to bound the search space for the perturbations ($\delta^\text{rx}_l,\delta^\text{tx}_l$). Rather than searching the entire space, the search space for ($\delta^\text{rx}_l,\delta^\text{tx}_l$) can be reduced to the grid area of the selected grid point.

\subsection{Finding Perturbations via Perturbation Solver $\mathbb{S}$}
At each iteration $k$, provided the initial grid points, the optimization problem in (\ref{PPOMP_DCOMP_Opt_Problem}) reduces to solving jointly for the $k$ perturbed parameters of the MPCs AoA-AoD and the cross-term gains $\Gamma_{l,q}, \forall l, q\in [k]$. The procedure to obtain these steps are detailed in Algorithm \ref{Algorithm:PDCOMP_Solver}. At this point, some remark on Algorithm \ref{Algorithm:PDCOMP_Solver} are in order
\begin{remark}
Due to the Hermitian structure, the cross-terms $\Gamma_{l,q},\forall l \in [k],\forall q \in [k]$ are only evaluated for $q\geq l$ terms (step 2 through 6). The terms $\Gamma_{l,q}, q< l = \Gamma_{l,q}^\text{*}$, thus saving the computational complexity exploiting the inherent Hermitian property of the covariance matrix.
\end{remark}

\begin{remark} At each iteration $k$, the AoA-AoD parameters are perturbed within their grid regions towards the direction that reduces the norm of the residual measurement covariance the most (step 8 in Algorithm 4). At the $p^\text{th}$ perturbation iteration, the AoA/AoD parameters are perturbed as $\theta^\text{tx}_{l,p} =  \bar{\theta}^\text{tx}_{l} + \delta^\text{tx}_{l,p}$ and $\theta^\text{tx}_{l,p} =  \bar{\theta}^\text{tx}_{l} + \delta^\text{tx}_{l,p}$, where $p$ is the perturbation index.
\end{remark}

At each perturbed point, the weights $\bm{\Gamma}$ and the perturbations will be updated sequentially in an alternating fashion as shown in steps 2 through 12 of Algorithm \ref{Algorithm:PDCOMP_Solver}. The matrices ${\mathbf{B}^\text{rx}} \in \mathcal{C}^{k \times (M_\text{RF}N_\text{RF} )^2}$ and ${\mathbf{B}^\text{tx}}\in \mathcal{C}^{k \times (M_\text{RF}N_\text{RF} )^2}$ holding the weighted partial derivatives with respect to the AoA and AoD, respectively, are mathematically defined as follows:
\begin{eqnarray}\label{B_Update_DCOMP}
\begin{aligned}
&\scalebox{1}{$ \mathbf{b}_l^\text{rx} =\left[  \left( \sum_{q=1}^k \Gamma_{l,q}\right) \text{vec}\left(\Phi\frac{\partial \mathbf{A}_\text{res}([\theta^\text{rx}_{l},\theta^\text{tx}_{l}],[\theta^\text{rx}_{q},\theta^\text{tx}_{q}])}{\partial \theta^\text{rx}_{1}}\Phi^\text{H}\right)\right],$}\\
&\scalebox{1}{$ \mathbf{b}_l^\text{tx} =\left[  \left(\sum_{q=1}^k \Gamma_{l,q}\right) \text{vec}\left(\Phi\frac{\partial \mathbf{A}_\text{res}([\theta^\text{rx}_{l},\theta^\text{tx}_{l}],[\theta^\text{rx}_{q},\theta^\text{tx}_{q}])}{\partial \theta^\text{tx}_{1}}\Phi^\text{H}\right)\right],$}\\
&\scalebox{1}{$\mathbf{B}^\text{rx} = \left[ \mathbf{b}_1^\text{rx}, \ldots,\mathbf{b}_k^\text{rx}\right]; \quad \mathbf{B}^\text{tx} = \left[ \mathbf{b}_1^\text{tx}, \ldots,\mathbf{b}_k^\text{tx}\right].$}
\end{aligned}
\end{eqnarray}

\begin{algorithm}[!b]
\KwIn{${\mathbf{R}_{{y}}}$, $\mathcal{S}_k$\\
\textbf{Initialization: } ${\mathbf{\Phi}}$, ${\mathbf{\Psi}}$, $p=1$, Initial grid points (from ${\mathcal{S}_k}$): $\theta^\text{rx}_{l,p} = \theta^\text{rx}_{l}$, $\forall l \in [k]$, $\theta^\text{tx}_{l,p} = \theta^\text{tx}_{l}$, $\forall l \in [k]$ }
\While{(Until the stopping criterion is met)}{
{
\For{l}{
\For{$q\geq l$}{
\scalebox{.87}{$\Gamma_{l,q} =  (\mathbf{\Phi} \mathbf{a}_\text{res}(\theta^\text{rx}_{l,p},\theta^\text{tx}_{l,p}) )^\dagger {\mathbf{R}_{{y}}} \left( (\mathbf{\Phi} \mathbf{a}_\text{res}(\theta^\text{rx}_{q,p},\theta^\text{tx}_{q,p}) )^\dagger\right)^\text{H}$}
}
}
$\Gamma_{l,q} = \Gamma_{q,l}^*, \quad \forall q < l$\\
$\mathbf{R}_{{y}_{\perp,p}}= \mathbf{R}_{y} - \Psi \sum_{l=1}^{k}\sum_{q=1}^{k} \Gamma_{l,q} \mathbf{A}(\Theta_l,\Theta_q) $,\\
\text{Update $\mathbf{B}^\text{rx}$ and $\mathbf{B}^\text{tx}$ as in (\ref{B_Update_DCOMP})}\\
\scalebox{.7}{${\theta^\text{rx}_{l,p+1} = \max{\{\bar{\theta}_l^\text{rx} -\Delta_\text{LB}^\text{rx},\min \{\bar{\theta}_l^\text{rx} +\Delta_\text{UB}^\text{rx}, \theta^\text{rx}_{l,p} +  \mu_{p} \mathbb{R}\{ \mathbf{B}_{(l,:)}^\text{rx} \text{vec}(\mathbf{R}_{{y}_{\perp,p}}) \}\} }\} }$}\\
\scalebox{.7}{${\theta^\text{tx}_{l,p+1} = \max{ \{ \bar{\theta}^\text{tx}_l - \Delta_\text{LB}^\text{tx},\min \{  \bar{\theta}^\text{tx}_l +\Delta_\text{UB}^\text{tx}, \theta^\text{tx}_{l,p} +  \mu_{p} \mathbb{R}\{ \mathbf{B}_{(l,:)}^\text{tx} \text{vec}(\mathbf{R}_{{y}_{\perp,p}}) \}\} }\} }$}\\
\scalebox{.85}{$\delta^\text{rx}_l = \theta^\text{rx}_{l,p+1}- \theta^\text{rx}_{l,p} \quad \forall l \in [k]; \quad \delta^\text{tx}_l = \theta^\text{tx}_{l,p+1}- \theta^\text{tx}_{l,p}\quad \forall l \in [k]$}\\
$p$ = $p$ + 1
}
}
\KwOut{\scalebox{.85}{$\bm{\Gamma} =[\Gamma_{1,1}, \ldots, \Gamma_{k,k}], \bm{\delta}^\text{tx} =[\delta_{1}^\text{tx}, \ldots, \delta_{k}^\text{tx}], \bm{\delta}^\text{rx} =[\delta_{1}^\text{rx}, \ldots, \delta_{k}^\text{rx}]$}}
\caption{Covariance Estimation: PPCOMP - Perturbation Solver $\mathbb{S}$} \label{Algorithm:PDCOMP_Solver}
\end{algorithm}

\section{Simulation Results}
In order to verify the efficacy of our proposed method, we consider a communications environment with $M \,{=}\, 16$ and $N \,{=}\, 8$ antennas at the BS and UE, respectively, and $K_L \,{=}\, 4$ MPCs with the number of clusters $K = 2$ and the number of MPCs per each cluster $L = 2$. The AoA-AoD tuple centers $\theta_{i}^\text{rx}$ and $\theta_{i}^\text{tx}$ are chosen randomly in the interval of $[0,\pi]$. Further, the complex gain $\alpha_{k,l,t}$ are modeled as $i.i.d$ random variables with the complex Gaussian distribution, $\alpha_{k,l,t} \sim \mathcal{CN}(0,1)$.

The covariance estimation algorithms are mainly evaluated based on the \textit{relative efficiency metric} as adopted in \cite{Relative_Metric_Eta_Caire}, which is defined as $\eta = \frac{\mathbf{U}_{\hat{\mathbf{R}}_{h}}^\text{H} \mathbf{R}_h \mathbf{U}_{\hat{\mathbf{R}}_h}}{\mathbf{U}_{\mathbf{R}_h}^\text{H} \mathbf{R}_h \mathbf{U}_{\mathbf{R}_h}} \in [0,1]$. Here $\mathbf{R}_h$ and $\hat{\mathbf{R}}_h$ are the true covariance and the estimated covariance matrix, respectively, while, $\mathbf{U}_{\mathbf{R}_h}$ and $\mathbf{U}_{\hat{\mathbf{R}}_h}$ are the matrices containing the singular vectors corresponding to the singular values of the true covariance and estimated covariance matrices, respectively. Intuitively, $1 -\eta$ denotes the fraction of signal power lost due to the mismatch between the optimal beamformer and its estimate \cite{Relative_Metric_Eta_Caire}. Thus, higher the $\eta$, better are the obtained estimates. We compare the performance of our proposed algorithms against the benchmark algorithm covariance OMP (COMP) proposed in \cite{park2018spatial}. All the results obtained are averaged over 100 independent trials.

\begin{figure}[t!]
    \centerline{
    \includegraphics[scale=0.7]{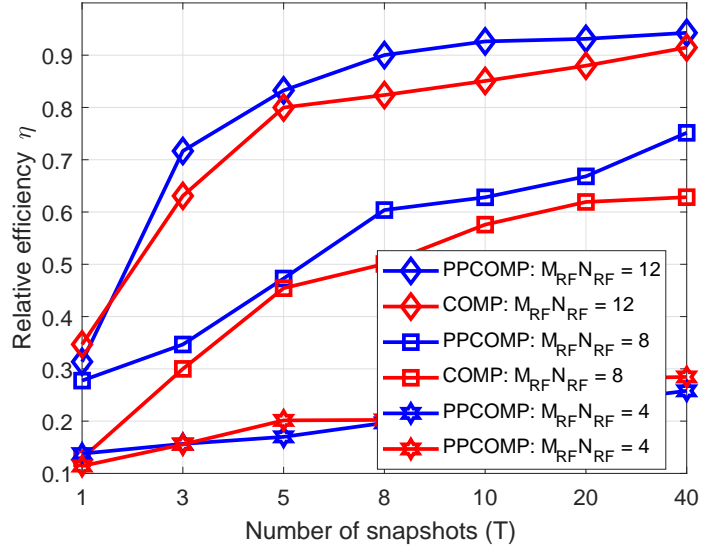}}
    \caption{Relative efficiency $\eta$ versus number of snapshots $T$, where the number of measurements is $M_{\rm RF}N_{\rm RF} \,{=}\, \{4, 8 ,12\}$, and the signal-to-noise ratio (SNR) is $10$ dB.}
    \label{fig:Eta_Vs_Snap}
\end{figure}

Fig. \ref{fig:Eta_Vs_Snap} compares the performance of COMP and PPCOMP in terms of relative efficiency $\eta$ with $M_\text{RF}N_\text{RF}$ = $\{4, 8, 12\}$ and SNR = 10 dB against the number of snapshots $T$. For the $M_\text{RF}N_\text{RF} \geq K_L$ regime\footnote{Due to the off-grid effects the number of significant non-zero elements in the sparse domain can be greater $K_L$.}, it can be seen that the PPCOMP outperforms COMP and is due to the fact that it is better equipped to capture the off-grid by means of controlled perturbed mechanism, whereas the COMP fails to do so. It is also observed that PPCOMP algorithm reach their peak performance at a smaller number of snapshots, which reduces the estimation time for fast changing environments in mmWave applications. On the other hand, the counterpart algorithm require relatively more snapshots to reach its peak performance which is lower than the perturbed version. Whereas for the $M_\text{RF}N_\text{RF} \leq K_L$ regime ($M_\text{RF}N_\text{RF} = 4$), both COMP and PPCOMP performs poor because the number of measurements are lesser than the number of MPCs (the number of non-zero elements in the sparse domain).

\begin{figure}[!t]
    \centerline{
    \includegraphics[scale=0.7]{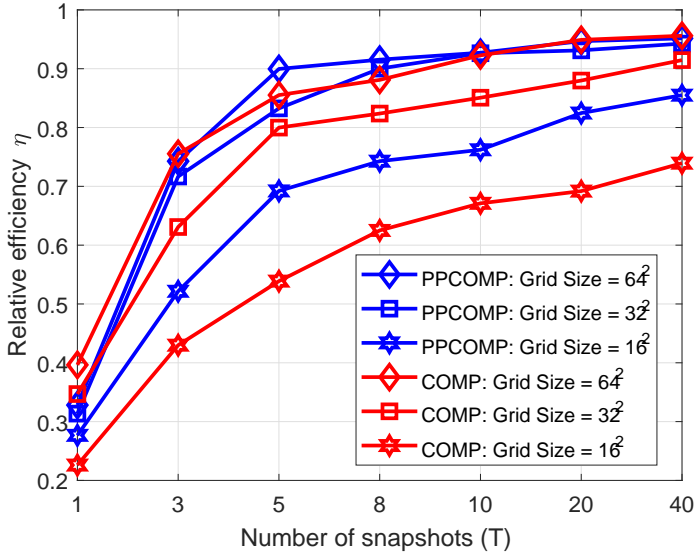}}
    \caption{Relative efficiency $\eta$ versus number of snapshots $T$ with different grid size. The number of measurements is $M_{\rm RF}N_{\rm RF} \,{=}\, 12$, and the SNR is $10$ dB.}
    \label{fig:Err_Vs_Snap}
\end{figure}
Fig. \ref{fig:Err_Vs_Snap} investigate the effect of number of grid points on the algorithms performance. The number of measurements was fixed to $M_\text{RF}N_\text{RF}$ = 12, SNR = 10 dB, and uniform sampling of $\cos(\theta)$ domain. It can be observed from Fig. \ref{fig:Err_Vs_Snap} that increasing the number of grid points (the level of discretization) can increase the performance of the COMP algorithm as it reduces the error caused due to the basis mismatch. Even though increasing the number of grid points has a positive effect, it also has negative effects. It increases the mutual correlation of the dictionary matrix and also leads to the undesirable increase in the computational complexity. To conclude, rather than using COMP over a larger and denser dictionary, it is advisable to use PPCOMP over a much smaller size dictionary \cite{teke2013perturbed}. For example, the PPCOMP with Grid size $= 32^2$ provides a similar performance as COMP with Grid size $= 64^2$ with lesser computational complexity.

We also evaluated the effect of SNR on covariance estimation. We noticed that at low medium SNR levels, the PPCOMP method is comparable to that of the COMP method. However,beyond medium SNR levels, the performance gap is significant.We also noticed PPCOMP provides better performance in terms of the normalized mean square error (reconstruction error) than the COMP version. However, due to space constraints, we omit its result.

\section{Conclusion}\label{sec:conclusion}
In this paper, we study the covariance estimation problem for MIMO mmWave network setup considering the off-grid effects. We propose the PPCOMP algorithms for the explicit covariance estimation. The proposed algorithms evade the issue arising from the basis mismatch problems by operating on the continuum AoA-AoD space using the mechanism of the controlled perturbation in conjunction with a modified OMP framework. The modified OMP framework helps to preserve the low computational complexity which is inherent for a greedy solver. On the other hand, the controlled perturbation mechanism jointly solves for the off-grid parameters and weights. Simulation results demonstrate the superiority of our proposed methods, and  outperforms the existing techniques both in terms of the relative efficiency metric.

\bibliographystyle{IEEEtran}
\bibliography{IEEEabrv,Bib_File}

\end{document}